# Observation of a Core-Excited Dipole-Bound State ~1 eV above the Electron Detachment Threshold in Cryogenically Cooled Acetylacetonate

Rafael A. Jara-Toro,[a,b,c,‡] Martín I. Taccone,[d] Jordan Dezalay,[e] Jennifer A. Noble,[e] Gert von Helden,[d] and Gustavo A. Pino,[a,b,c,*]

a- INFIQC: Instituto de Investigaciones en Fisicoquímica de Córdoba (CONICET – UNC) - Haya de la Torre y Medina Allende, Ciudad Universitaria, X5000HUA Córdoba, Argentina.

b- Departamento de Fisicoquímica, Facultad de Ciencias Químicas – Universidad Nacional de Córdoba – Haya de la Torre y Medina Allende, Ciudad Universitaria, X5000HUA Córdoba, Argentina.

c- Centro Láser de Ciencias Moleculares - Universidad Nacional de Córdoba - Haya de la Torre s/n, Pabellón Argentina, Ciudad Universitaria, X5000HUA Córdoba, Argentina.

d- Fritz Haber Institute of the Max Planck Society, 14195 Berlin, Germany

e- Physique des Interactions Ioniques et Moléculaires (PIIM): CNRS / Aix-Marseille Université UMR 7345, Avenue Escadrille Normandie-Niémen, 13013 Marseille, France.

Rafael A. Jara-Toro: 0000-0002-3837-5161
Jennifer A. Noble: 0000-0003-4985-8254
Martín I. Taccone: 0000-0002-6795-2693
Jordan Dezalay: 0009-0006-9156-4718
Gert von Helden: 0000-0001-7611-8740
Gustavo A. Pino: 0000-0002-8702-2688

* Corresponding author: gpino@unc.edu.ar

‡ Present address: IPR (Institute de Physique de Rennes) - Université de Rennes, CNRS, UMR 6251, F-35000 Rennes, France

**ABSTRACT**

Dipole-Bound States in anions exist when a polar neutral core binds an electron in a diffuse orbital through charge-dipole interaction. Electronically excited polar neutral cores can also bind an electron in a diffuse orbital to form Core-Excited Dipole-Bound States (CE-DBS), which are difficult to observe because they usually lie above the electron detachment threshold leading to very short lifetimes and, thus, unstructured transitions. We report here the photodetachment spectroscopy (PDS) of cryogenically cooled acetylacetonate anion ($C_5H_7O_2^-$) recorded by detecting the neutral radical produced upon photodetachment and the IR spectroscopy in He-nanodroplets. Two DBSs were identified in this anion. One of them lies close to the electron detachment threshold (∼ 2.74 eV) and is associated with the ground state of the radical ($D_0$-DBS). Surprisingly, the other DBS appears as resonant transitions at 3.69 eV and is assigned to the CE-DBS associated with the first excited state of the radical ($D_1$-DBS). It is proposed that the resonant transitions of the $D_1$-DBS are observed ∼1 eV above the detachment threshold because its lifetime is determined by the internal conversion (IC) to the $D_0$-DBS, after which the fast electron detachment takes place.

# I. INTRODUCTION

As in the case of Rydberg states in neutral molecules, valence-bound states (VBSs) of anions with polar neutral cores with a large enough dipole moment ($\mu \geq 2.5$ D) can support diffuse dipole-bound states (DBSs) below the detachment threshold, due to the long-range electron-dipole interaction.[1] The DBSs were first observed in photodetachment spectroscopy (PDS) of anions to obtain electron affinities of organic radicals.[2-8]

Even though the critical dipole moment required to support a DBS is $\mu \geq 2.5$ D, it was shown that the polarizability of the neutral core is also at play and polarization-assisted-DBSs in species with $\mu < 2.5$ D have been reported recently.[9-11]

Anions usually do not possess excited valence-bound states (E-VBSs) because of their relatively low electron binding energies. However, non-valence DBSs constitute a special type of excited electronic state of the anion and for this reason they are also called excited DBSs, even when the electron is bound to the ground state neutral-core.

Electronically excited states of neutral polar molecules or radicals with a sufficiently large $\mu$ can also bind an electron to form core-excited dipole-bound states (CE-DBEs). However, since the dipole-bound electron is a quasi-free electron, CE-DBSs are very difficult to observe since they generally lie above the electron detachment energy. Although CE-DBSs have been implicated in a few previous works,[12-15] to the best of our knowledge there exist only three reports of their experimental observation; in the potassium iodide KI$^-$ anion,[16] the pyrazolide anion[17] and the nitromethane anion.[18]

The dynamics of excited DBSs produced upon excitation to E-VBSs of anions has also been extensively studied.[19-21] Very recently, it was demonstrated that the dissociation of iodophenoxide anions along the C-I coordinate is mediated by Feshbach resonances of DBSs, playing a critical role as the doorway into the anion´s chemistry.[22]

Understanding the spectroscopy and dynamics of DBSs is important because they have many implications in physical and chemical processes of relevance in different areas of science, including physics, chemistry, biology, and astronomy. For instance, DBSs have been suggested to be involved in the formation of anions with VBSs though electron attachment (EA) or even dissociative electron attachment (DEA). In this regard, they are potentially involved in biomolecule damage by low-energy DEAs, including DNA[23-25] and proteins,[13,15] as well as in the formation of anions in the interstellar medium[5,26] and have also been proposed as possible carriers of the diffuse interstellar bands (DIBs).[27,28]

The development of very sophisticated techniques such as pump-probe spectroscopies and velocity mapping imaging (VMI) to study electron detachment processes coupled with methodologies to cool down the ions (cryostatic cooling, cryo-SEVI, etc.) have allowed access to a deeper knowledge on the properties of DBSs. A complete review on the recent advances on DBSs can be found in Refs. 7, 8, 21, 29 and 30.

In recent years, we have been studying the PDS of cryostatically cooled aromatic anions by detection of the neutral radical or the neutral fragments as a function of the photon excitation energy.[31-33] As a continuation of these studies to a non-aromatic, conjugated system, we present here the study of the PDS of the acetylacetonate ($C_5H_7O_2^-$) anion.

Aceylacetonate is a β-diketonate type of chelating ligand employed to produce volatile transition metal complexes used for chemical vapor deposition.[34,35]

A study by Irikura[36] made a summary of the available information and predicted the structure, vibrational spectra and electron detachment energies of the $C_5H_7O_2^-$ anion. Ab-initio calculations at the MP2/6-311G**//HF/6-311G* level have shown that the $C_5H_7O_2^-$ anion has three stable di-keto structures (Scheme 1) lying within 0.28 eV.[36] The most stable is a *trans* isomer (aT) with the two C=O groups in *anti*-conformation, followed by two *cis* isomers (aC1 and aC2) with the carbonyl groups pointing *exo* and *endo,* respectively. Although aC2 is the least stable, this is the metal chelating structure of all the transition metal complexes. Hereafter the anions will be called aX and the corresponding radicals rX (X = T, C1 or C2, depending on the isomer).

aT         aC1         aC2

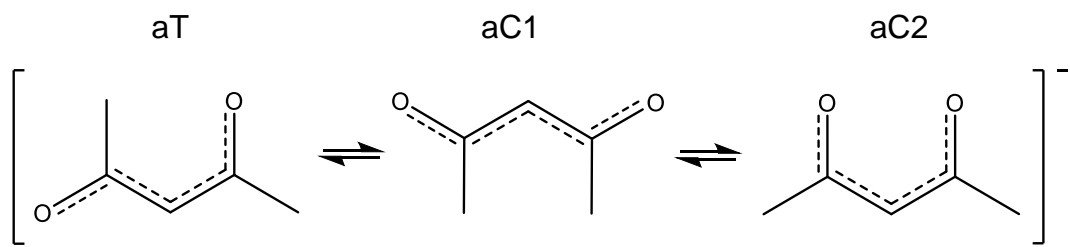

**Scheme 1:** *The three stable di-keto isomers of $C_5H_7O_2^-$ anion.*

The electron detachment energy of the acetylacetonate anion was experimentally determined by electron transfer bracketing to be 2.52 ± 0.22 eV,[34] while the adiabatic detachment energy (ADE) for the three isomers was determined at the B3LYP/6-311G**//HF/6-311G* level to be 2.8 eV.[36]

## II. METHODOLOGY

### A. Experimental set-up in Aix-Marseille for UV-vis spectroscopy

Acetylacetonate ($C_5H_7O_2^-$) anions are produced in an electrospray ionization (ESI) source by injecting a solution of zinc acetylacetonate ($Zn(C_5H_7O_2)_2$) (99.9% – Sigma-Aldrich Merck – Darmstadt, Germany) 60 μM in 1:1 mixture of methanol:water.

The experimental setup for cold ion photofragmentation spectroscopy, described in previous publications,[37] has been modified to allow for the detection of both neutral and negative photoproducts of negative ions.[31-33]

The modified setup has been detailed elsewhere[31-33] and only a brief outline will be given here. Ions from the ESI source are guided into the cooled quadrupole-ion-trap (QIT) just after a pulse of helium gas. The ions are stored in the cold QIT for a few tens of ms, the time necessary for cooling to c.a. 50 K via collisions, as well as the decrease of the He pressure in the trap via pumping. The cooled ions are then extracted before being accelerated to 2.6 kV, with the extraction and acceleration voltages chosen to optimize focusing on the detector following Wiley McLaren.[21] Immediately after acceleration, the ions enter the "Gauss tube" which is initially set to the accelerating potential.

In the experiments presented here, the laser interacted with the parent anions in the Gauss tube. To detect the parent ions, the voltage on the Gauss tube was set to 0 V once the ions entered. In the neutral detection mode, to prevent any unreacted parent anions from reaching the detector and complicating the detection of the neutral photodetached radical, the voltage on the Gauss tube was left on in order to deviate the anions. The neutrals are not perturbed by the voltage.

Since the parent anions already had kinetic energy due to their accelleration, any neutral parent radicals produced via photodetachment and/or any neutral daughter fragments resulting from photofragmentation continue to travel in the field-free region of the time of flight (TOF) mass and are detected on the MCP. While the detection of narrow peaks, corresponding to neutral radicals produced by photodetachment of the parent anion, is the only TOF signature relevant to the present study, in some other molecular systems we have observed widened temporal profiles related to fragmentation (see Ref. 33 for a resume of the neutral detection method).

The ions were photoexcited using a tunable OPO laser (EKSPLA) which operates at a repetition rate of 10 Hz, and has a 10 ns pulse width with ~ 10 $cm^{-1}$ spectral linewidth). Each ion bunch interacts with only one laser pulse, with the measured signal typically averaged over at least eight shots per wavelength, with the wavelength tuned automatically using an in-house interface. The spectra of neutrals are averaged over around ten total measurements, while no correction is made for the minor fluctuations of the laser power during tuning.

**B. Experimental set-up in Berlin for IR spectroscopy**

The experimental setup for cold-ion infrared action spectroscopy in helium droplets has been described previously. Briefly, gas phase ions are generated with a nanoelectrospray ionization (nESI) source using Pd/Pt-coated borosilicate capillaries fabricated in-house. Samples of 0.1% acetylacetone (99.9% – Sigma-Aldrich Merck – Darmstadt, Germany) or 100 µM $Zn(C_5H_7O_2)_2$ (99.9% – Sigma-Aldrich Merck – Darmstadt, Germany) in $ACN:H_2O$ 1:1 solutions (99.9% – Sigma-Aldrich Merck – Darmstadt, Germany) were employed. Once transferred into the setup, the ion mass of interest is isolated using a quadrupole mass filter before being deflected into a hexapole ion trap using a quadrupole ion bender at 90º. He buffer gas is introduced into the trap in a 2.0 s pulse just prior to the arrival of the ions to collisionally cool and aid in trapping the ions. The ion trap is cooled down to ca. 90 K by a liquid $N_2$ flow, which effectively reduces the residual water pressure and potential ion-molecule reactions in the trap. After pumping for a further 1.5 s, helium droplets (generated via an Even–Lavie pulsed valve operated at 19 K with a helium backing pressure of ca. 70 bar) traverse the trap. . During their passage through the trap, He droplets can pick up an ion and, due to their high kinetic energy, overcome the longitudinal trap potential. Previous experiments under similar conditions have indicated that these ion-doped He droplets consist of ca. 20000 He atoms. After exiting the ion trap, doped droplets are irradiated with infrared photons of the Fritz-Haber-Institut Free-Electron Laser (FHI-FEL) in the '*interaction region*' of the setup. Photons that are resonant with vibrational transitions of the ion may be absorbed. This energy is rapidly dissipated via the evaporation of helium atoms and, following several cycles of photon absorption and energy dissipation, the completely bare ion photoproducts are detected with a TOF mass analyzer. By tuning the IR wavelength, infrared action spectra are obtained.

The FHI-FEL radiation consists of a series of laser macropulses with a duration of 10 μs, each of which contains micropulses of ca. 5 ps and energy of ca. 10 μJ, with the repetition rate set to 1 GHz. For each IR wavelength position, the ion trap is filled, before measuring TOF signals from 25 laser pulses at a frequency of 10 Hz. After repeating this process at all wavelengths, the IR spectrum obtained is then corrected by dividing the ion intensity by the photon fluence per wavelength. The final spectrum is the average of at least three individual scans.

## C. Calculations

The structure and stability of various isomers of the $C_5H_7O_2^-$ anion were calculated using density functional theory (DFT) with the CAM-B3LYP functional with the aug-cc-pVDZ basis set.[38] The Adiabatic and Vertical Detachment Energies (ADE and VDE, respectively) and the $S_n \leftarrow S_0$ Adiabatic and Vertical Transition Energies of the anions ($E_{ad}$ and $E_v$) were also calculated at the DFT and TD-DFT level with the same functional and basis set, as well as the Franck-Condon factors (FCF) simulation. By comparison with previous works on similar systems, the estimated energy uncertainty of this methodology is in the order of ± 0.2 eV for broad band unstructured spectra[31-33,39] which is reduced to ≤ 0.1 eV for structured transitions.[40] In this work, in addition to the DFT calculations, the ground state energy of the isomers was refined at the CCSD(T)/aug-cc-pVDZ//CAM-B3LYP/aug-cc-pVDZ level of theory, showing a correction to the DFT energy ≤ 10 %, as shown in Table I.

The ground state vibrational IR spectra were calculated using the B3LYP-D3(BJ)/aug-cc-pVTZ level of theory with a scaling factor of 0.98 for all the vibrational frequencies. All calculations were performed with the Gaussian 16 suite of programs.[41]

## III. RESULTS

### A. UV-vis photoelectron detachment spectrum

The electron photodetachment spectrum obtained by recording the intensity of the neutral parent $C_5H_7O_2^•$ radical produced by the interaction of the laser with the $C_5H_7O_2^-$ anion in the Gauss tube as a function of the photon energy in the 2.6-4.1 eV spectral range is shown in Figure 1. The spectrum is composed of a continuous background signal due to direct electron detachment governed by the Wigner law,[42] with a smooth onset at (2.74 ± 0.02) eV. In addition, two group of resonant peaks superimposed on the continuum were observed and are shown in the insets of

Figure 1, recorded with a resolution of 126 points/nm. These peaks indicate the presence of resonances, either due to excited anionic states, E-VBS ($S_1$, $S_2$, …), or to DBSs.

The first group of bands starts at 2.782 eV with a progression of 0.006 eV (48 cm$^{-1}$) and an average full-width half-maximum (FWHM ∼ 15 cm$^{-1}$). A second group of more intense bands appears at 3.691 eV with a FWHM = 35 cm$^{-1}$ at the band origin.

The onset at 2.74 eV and the step at 2.83 eV are higher than the previously reported value (2.52 ± 0.22 eV)[35] by electron capture at 350 K, and it could be related to the internal energy of the system at higher temperatures leading to lower measured values; and thus, our result are closer to the actual ADE value.

No fragmentation of the $C_5H_7O_2^-$ parent was observed in the spectral range (2.6 – 5.4) eV, nor was any fragmentation of the neutral $C_5H_7O_2^\bullet$ radical photoproduct observed, regardless of the photon energy used to photodetach the $C_5H_7O_2^-$ parent anion.

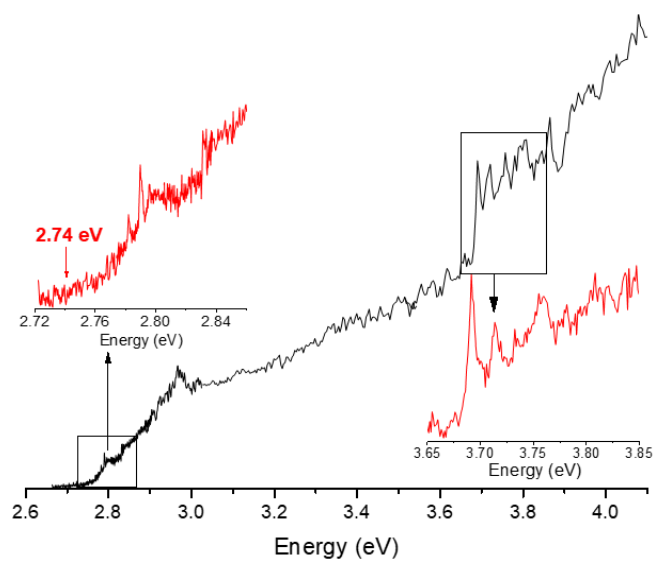

***Figure 1:*** *Electron photodetachment spectrum obtained by recording the intensity of the neutral parent $C_5H_7O_2^\bullet$ radical produced by the interaction of the laser with $C_5H_7O_2^-$ anion in the Gauss tube. The two insets (red lines) show higher resolution scans in the regions where the two groups of bands are observed.*

## B. IR spectrum in He-nanodroplets

IR spectra of $C_5H_7O_2^-$ produced from two different precursors (acetylacetone or $Zn(C_5H_7O_2)_2$) and cooled in He-nanodroplets are shown in Figure 2. The spectra show no differences irrespective of the $C_5H_7O_2^-$ anion precursor used.

The unsaturated IR spectrum measured at 85 % attenuation of the FEL and using acetylacetone as precursor (green) shows a very intense transition at 1550 cm$^{-1}$, a weak transition at 992 cm$^{-1}$ and several very weak transitions that are best observed at higher laser powers where the main band at 1550 cm$^{-1}$ is saturated.

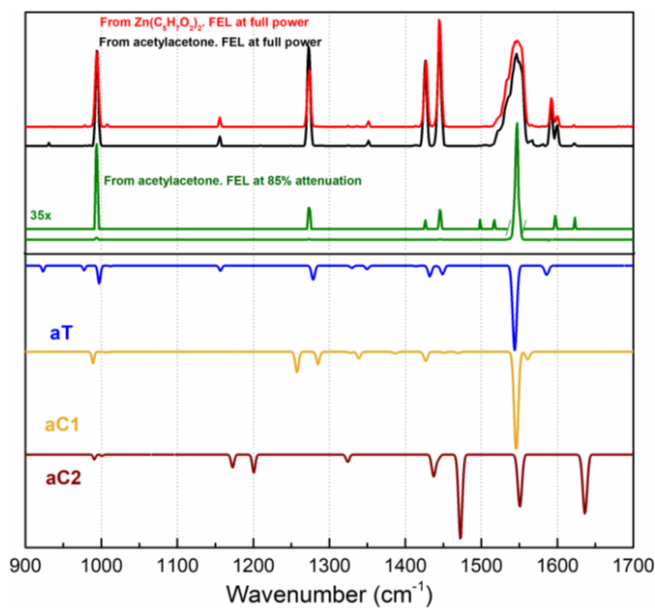

*Figure 2:* Cold-ion IR action spectra of $C_5H_7O_2^-$ anion inside He nanodroplets, generated from $Zn(C_5H_7O_2)_2$ (red) and pure acetylacetone (black) in the 900 - 1700 $cm^{-1}$ region. The green line shows the same spectrum obtained with 85% attenuation of the FEL using acetylacetone as precursor. A zoom in (x 25) of this spectrum is also shown in green. Theoretical spectra for the aT (blue), aC1 (yellow) and aC2 (brown) isomers were calculated at B3LYP-D3(BJ)/aug-cc-pVTZ level of theory with a scaling factor of 0.98.

### C. Theoretical

The ground state geometry optimization was performed for the three isomers of Acetylacetonate. The optimized structures and their relative electronic energies (EE), with and without the zero point energy correction (ZPE), the vertical detachment energy (VDE) and adiabatic detachment energy (ADE) were calculated and reported in Table I. The dipole moments ($\vec{\mu}$) of the optimized structures of the corresponding radicals in their ground state are also shown in Table I.

The calculated ADE+ZPE reported in Table I for all isomers are in the range of 2.74 - 2.68 eV, in very good agreement with the experimental electron photodetachment threshold (2.74 ± 0.02) eV observed in Figure 1. Due to their proximity, these values cannot be used to discriminate between the different isomers.

As shown in Table I, the ground state of the three radicals have dipole moments large enough ($\vec{\mu} \geq 2.5$ D) to support the DBSs observed in the electron photodetachment spectrum.

**Table I:** *Calculates relative energy of the four isomers of $C_5H_7O_2^-$ ($\Delta E$) with and without zero point energy correction (ZPE). Vertical and adiabatic electron detachment energies (ADEs and VDEs). The calculations were performed at the DFT level using CAM-B3LYP functional with the aug-cc-pVDZ basis set. Refined values at the CCSD(T)/aug-cc-pVDZ//CAM-B3LYP/aug-cc-pVDZ theory level are shown in parenthesis.*

| | Anion | | | | | | Radical |
|---|---|---|---|---|---|---|---|
| Isomers $S_0$ geometry | $\Delta E$ eV | $\Delta E$ + $\Delta$ZPE eV | $\Delta G°_{298K}$ gas phase eV | VDE eV | ADE + $\Delta$ZPE eV | $\vec{\mu}$ D | Isomers $D_0$ geometry |
| 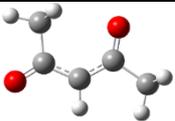 aT | 0.00 | 0.00 | 0.00 | 2.98 | 2.74 | 2.5 | 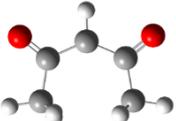 rT |
| 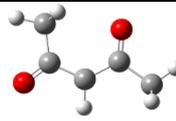 aC1 | 0.19 (0.17) | 0.19 (0.18) | 0.20 (0.18) | 2.91 | 2.69 | 3.8 | 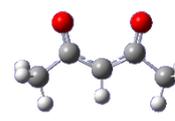 rC1 |
| 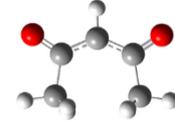 aC2 | 0.29 (0.29) | 0.29 (0.28) | 0.26 (0.26) | 2.90 | 2.68 | 5.4 | 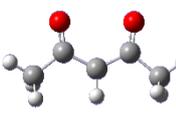 rC2 |

## IV. DISCUSSION

### A. Vibrational IR spectroscopy

The comparison of the experimental IR spectrum with the IR spectra calculated for the three isomers of the anion is shown in Figure 2.

Although the very intense band at 1550 cm$^{-1}$, assigned to the antisymmetric C=O stretching mode and the weak band at 992 cm$^{-1}$, assigned to the in-plane C-H bending mode are common features for the three isomers, all other bands can be confidently assigned to the aT isomer, which in turn is the most stable one (Table I).

Although there are a few bands that could be tentatively assigned to the aC1 and aC2 isomers, their relative intensities suggest that, in the case that they are present, their population is significantly lower than the population of the aT isomer, irrespectively of the precursor used to produce the $C_5H_7O_2^-$ anion. Therefore, hereafter we will consider only the aT isomer for the discussion.

### B. Electronic UV-vis spectroscopy

If we assume that only the aT isomer is populated at the low temperature of the experiment, as suggested by the IR spectrum, the calculated ADE for this isomer (2.74 eV) is in very good agreement with the smooth onset (2.74 eV) of the continuous signal.

As mentioned in section III.A, the sharp transitions could be due to DBSs or to E-VBS ($S_1$, $S_2$, …) and the comparison with the photoelectron spectrum (PES) recorded upon excitation of these resonances is needed to unambiguously assign them; but, we were unable to record the PES.

However, the adiabatic transition energy ($E_{ad}$) first E-VBS ($S_1$) is calculated at 3.30 eV, as shown in Table II. Considering the accuracy of these calculations as compare with other similar systems,[31-33,39]**Error! Bookmark not defined.** it is unlikely that the $S_1$ state is responsible for the observed resonances (2.782 eV – 2.794 eV).

On the other side, as shown in Table I, the ground state of the rT radical has a dipole moment large enough ($\vec{\mu}$ = 2.5 D) to support a DBS, that could be responsible for the three resonant transitions observed in this spectral region. However, if we assume that the onset at 2.74 eV corresponds to the ADE, the ground vibrational level of the DBS ($0_0^0$ transition) should be observed just below the detachment threshold, which is not the case.

To get more information, and considering that the electron in a DBS has little effect on the vibrational structure of the neutral core,[7,11] the FCF for the aT($S_0$) → rT($D_0$) transition were calculated and the corresponding vibronic spectrum is shown in Figure 3 in comparison with the experimental spectrum.

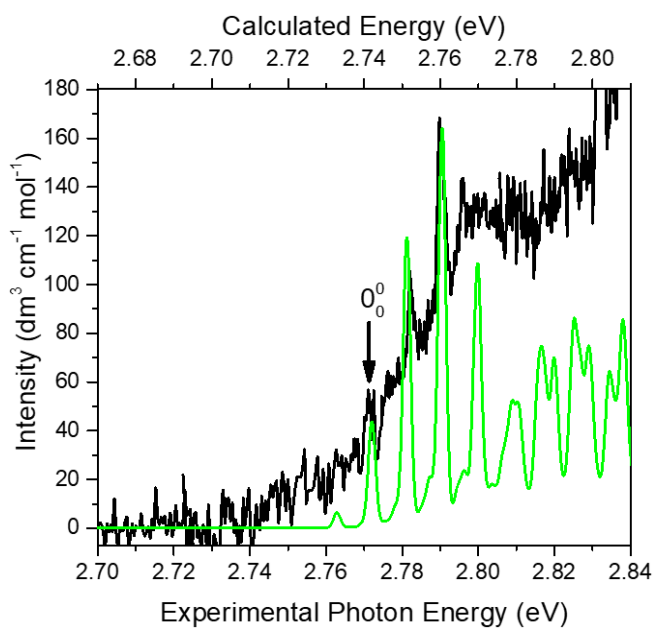

*Figure 3: Electron photodetachment spectrum recorded in the 2.7 eV region (black). Simulated spectrum for the $S_0 \rightarrow D_0$ transition of the aT isomer at 50 K.*

According to the FCF simulation shown in Figure 3, the $0_0^0$ transition for the DBS should appear at 2.77 eV, and the actual ADE is expected to be above this energy. From comparison with other systems containing neutral cores with $\vec{\mu}$ ~ 2.5 D, the binding energy of DBSs is expected to be < 10 cm$^{-1}$ (~1 meV),[10,39] which is below the spectral resolution of the experimental set-up (15 cm$^{-1}$ including the bandwidth of the laser and the rotational broadening at 50 K). It has been shown that the binding energy could increase up to ~ 0.001 eV for neutral cores with $\vec{\mu}$ ~ 2.5 D and with high quadrupole moment and high polarizability.[43] These limiting binding energies (0.001 eV – 0.01 eV) allow setting the ADE in the range 2.77 eV - 2.78 eV and thus, the three well-resolved transitions starting at 2.782 eV are assigned to excited vibrational levels of the DBS, also known as vibrational Feshbach resonances.

The $0_0^0$ transition for the DBS, which is usually very weak because it could only be observed by two-photon absorption or collision induced detachment, is not strong enough to be detected above the noise level given the overall signal of the anion, as previously reported for $C_2P^-$.[44]

As observed in the FCF simulation shown in Figure 3, at 50 K there are some hot bands appearing at lower energy than the $0_0^0$ transition. Therefore, the photodetachment onset observed at 2.74 eV could due to the photodetachment from hot anions.

Another possibility is to consider that the onset of the continuous signal at 2.74 eV is due to the presence of aC1 or aC2 isomers, in addition to the aT isomer, since their corresponding ADEs are calculated below the ADE of the aT isomer.

Although, we are not able to rule out either hypothesis, the IR spectroscopy strongly suggests the presence of the aT isomer only and it strengthens the first assumption. In any case, the PES would be needed for a precise determination of the ADE.

The bandwidth of these transitions is in the order of ~ 15 $cm^{-1}$, as is the bandwidth of the laser and the rotational contour at 50 K. Thus, the lifetime broadening is < 1 $cm^{-1}$ leading to an estimated lower limit lifetime > 5 ps. Considering the uncertainty of this estimation, it can only be concluded that the DBS of acetylacetonate has a lifetime longer than 5 ps, and it could be as long as a few nanoseconds.

This lifetime range (ns-ps) is supported by recent and accurate determinations of the lifetimes of various vibrational level of the DBS of phenoxide anion using a pump-probe experiment, which was reported to range from the nanosecond scale at the band origin,[45] to a few picoseconds for vibrational Feshbach resonances.[46]

While the sharp transitions in the 2.7 eV region are assigned to the DBS associated to the electronic ground state of the rT radical ($D_0$), the sharp transitions starting at 3.69 eV are more difficult to understand and their assignment is the main goal of this work.

Once more, these transitions could be assigned to shape resonances due to above threshold excitation to low lying E-VBS ($S_1$, $S_2$, …) or to CE-DBS ($D_1$, $D_2$…). In this regard, the vertical ($E_v$) and $E_{ad}$ transition energy to the first five excited states of aT and of the first excited state of rT were calculated and the results are shown in Table II and in Figure 4. The information for the $S_2$ excited state of aT is not reported because its oscillator strength (O.S.) is null.

As shown in Figure 4, the electronic configurations of the $S_0$, $S_1$, $S_3$, $S_4$ and $S_5$ states of aT are $S_0(…n^2, \pi^2)$, $S_{1,3,4}(…n^2, \pi^1, \sigma_n^{*1})$, with different $\sigma$ orbitals, and $S_5(…n^1, \pi^2, \sigma_n^{*1})$, respectively, while those of the $D_0$ and $D_1$ states of rT are $D_0(…n^2, \pi^1)$ and $D_1(…n^1, \pi^2)$, respectively. According to their electronic configurations, the $S_1$, $S_3$ and $S_4$ excited states of aT adiabatically correlate with the $D_0$ state of rT, while the $S_5$ state of aT correlates with the $D_1$ state of rT.

The calculated $E_{ad}+\Delta ZPE$ for the $S_0 \rightarrow S_3$ (3.82 eV) and $S_0 \rightarrow S_4$ (3.85 eV) are close to the experimental value (3.69 eV). However, these states correlate directly with the $D_0$ state of the radical (2.74 eV) (Figure 4), leading to an excess energy of around 1.1 eV in the $D_0$ state that would induce a very fast direct electron detachment and thus, broad bands (FWHM > 200 cm$^{-1}$) are expected for these transitions as usually observed for shape resonances.[11]

Therefore, the possibility of assigning the transitions observed in the 3.69 eV region to shape resonances is dismissed and they are more likely assigned to transitions to a CE-DBS as was recently reported for KI$^-$,[16] pyrazolide,[17] and CH$_3$NO$_2^-$ [18] anions.

In the cases of of pyrazolide[16] and KI$^-$[17] the $0_0^0$ transitions of the DBS and CE-DBSs are all located below the ADE, at variance with aT, in which the $0_0^0$ transitions is found 0.92 eV above the estimated ADE as in the case of CH$_3$NO$_2^-$.[18]

**Table II:** *Vertical and adiabatic electronic transition energies ($E_v$ and $E_{ad}$) for the $S_0 \rightarrow S_n$ transitions of the aT isomer and $D_0 \rightarrow D_1$ transition of the rT isomer, with ZPE correction of both states ($\Delta$ZPE) and the corresponding oscillator strengths (O.S.). The calculations were performed at the TD-DFT level using CAM-B3LYP functional with the aug-cc-pVDZ basis set.*

| Anion (aT) | | | | | | | |
|---|---|---|---|---|---|---|---|
| $S_0 \rightarrow S_1$ | | $S_0 \rightarrow S_3$ | | $S_0 \rightarrow S_4$ | | $S_0 \rightarrow S_5$ | |
| $E_v$ eV | $E_{ad}+\Delta$ZPE eV | $E_v$ eV | $E_{ad}+\Delta$ZPE eV | $E_v$ eV | $E_{ad}+\Delta$ZPE eV | $E_v$ eV | $E_{ad}+\Delta$ZPE eV |
| 3.53 | 3.30 | 4.13 | 3.82 | 4.13 | 3.85 | 4.43 | 4.33 |
| O.S.: 2 x 10$^{-3}$ | | O.S.: 2.7 x 10$^{-2}$ | | O.S.: 6 x 10$^{-3}$ | | O.S.: 8 x 10$^{-2}$ | |
| 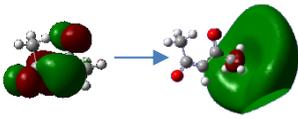 | | 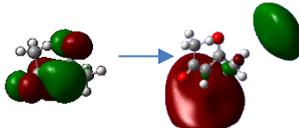 | | 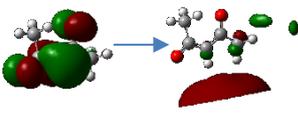 | | 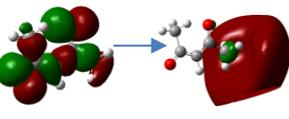 | |
| HOMO | LUMO | HOMO | LUMO + 1 | HOMO | LUMO + 2 | HOMO - 1 | LUMO |
| Radical (rT) | | | | | | | |
| $S_0 \rightarrow D_0$ | | | $D_0 \rightarrow D_1$ | | | | |
| VDE eV | ADE+$\Delta$ZPE eV | $\vec{\mu}(D_0)$ D | $E_v$ eV | $E_{ad}+\Delta$ZPE eV | $\vec{\mu}(D_1)$ D | | |
| 2.98 | 2.74 | 2.5 | 0.96 | 0.94 | 4.4 | | |
| SOMO (D$_0$) 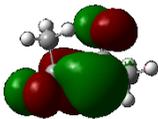 | | | SOMO (D$_1$) 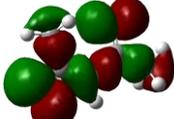 | | | | |
| $S_0 \rightarrow D_1$ $E_{ad}+\Delta$ZPE (eV) = 3.68 eV | | | | | | | |

The $E_{ad}+\Delta$ZPE for the first electronically excited state (D$_1$) of rT was calculated at 0.94 eV above the D$_0$ state, which is 3.68 eV above the S$_0$ state of the aT anion, in very good agreement with the experimental value of 3.69 eV determined for the $0_0^0$ transition of this group of resonances (Table II).

The enlarged $\vec{\mu}(D_1)$ = 4.4 D of rT in the $D_1$ state as compared with that of the $D_0$ state ($\vec{\mu}(D_0)$=2.5 D) allows for the rT($D_1$) to support a CE-DBS and the binding energy of the electron to the $D_1$ excited-core is expected to be larger than to the $D_0$ ground state-core. Hereafter, the DBS and CE-DBS will be called $D_0$-DBS and $D_1$-DBS, respectively, to label the electronic state of rT to which the electron is bound.

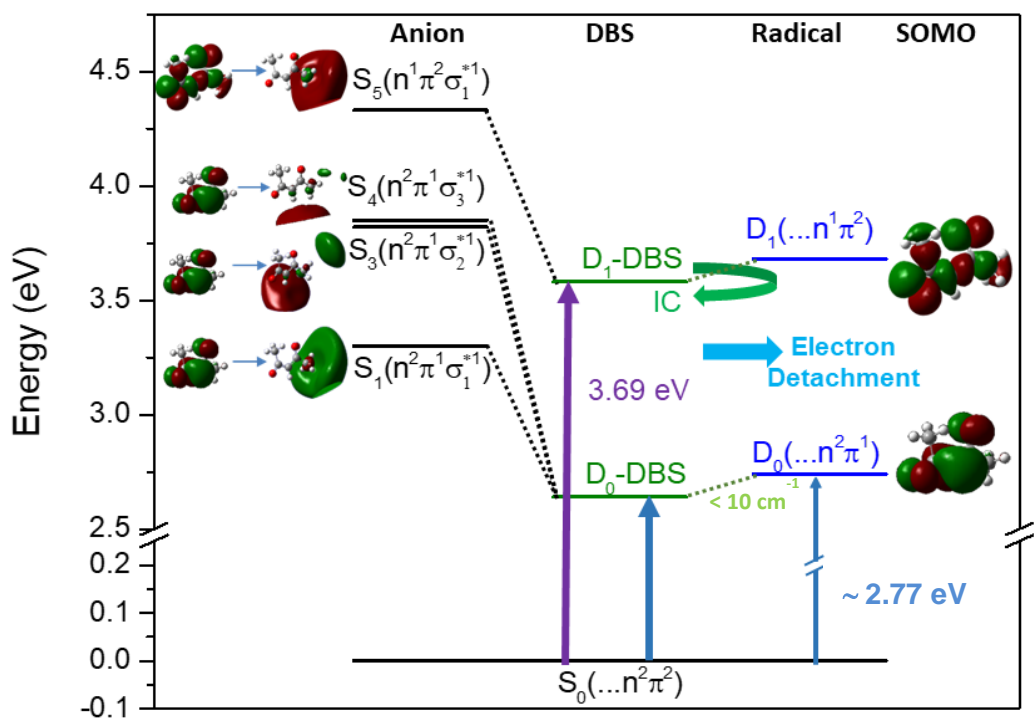

*Figure 4:* Schematic energy level diagram showing the photodetachment from the $S_0$ state of aT. The two-electron photodetachment process through the $0_0^0$ transition of the $D_1$-DBS of rT is also shown. On the right part of the figure, the SOMOs corresponding to $D_0$ and $D_1$ states are shown. They correspond to the HOMO and HOMO-1 of the anion.

According to their electronic configurations, the $D_0$-DBS correlates with the $D_0$ state of rT, which corresponds to detaching an electron from the $\pi$ orbital (HOMO) of aT($S_0$). On the other hand, the $D_1$-DBS correlates with the $D_1$ state of rT, which corresponds to the removal of an electron from the $n$ orbital (HOMO - 1) of aT($S_0$) (see Table II).

Under the present experimental conditions, we are able to determine neither the electron affinity of the electronically excited radical nor the binding energy of the electron to the $D_1$ state, for which the photoelectron spectrum would be required. Whatever the binding energy is, at the band origin of the $S_0 \rightarrow D_1$-DBS there is not enough energy to produce the electron detachment concurrently with electronically excited rT($D_1$) (Figure 4). However, the electron detachment can still occur if the system reaches the $D_0$-DBS through internal conversion (IC).

The lifetime of the $D_1$-DBS, and the bandwidth of the $0_0^0$ transition, will thus be determined mainly by the rate of the IC to the $D_0$-DBS, which requires the relaxation of an electron from the $\pi^2$ orbital to the $n^1$ orbital ($D_1$-DBS ($n^1$, $\pi^2$) → $D_0$-DBS ($n^2$, $\pi^1$)). So that, the overall electron detachment process from the ground state of aT($S_0$) through the $0_0^0$ transition of the $D_1$-DBS is a two-electron process and consequently, it is slower than the direct electron detachment.

Narrow resonances assigned to a slow two-electron process leading to the detachment from Feshbach states have been previously reported for the *p*-benzoquinone radical anion.[47] In that case, the neutral core is a closed-shell non-polar molecule and the authors suggested that the enhanced photodetachment takes place via resonantly excited anion states. The broadening of the transitions allowed them to estimate the lifetime of the two-electron detachment in the range of 0.2 and 1.2 ps.

In the case of aT, the width for the $0_0^0$ band of the $D_1$-DBS is 35 cm$^{-1}$ while the broadening due to the rotational contour at 50 K and the laser bandwidth is ~15 cm$^{-1}$ according to the simulation. So that, the lifetime broadening of this band is ~20 cm$^{-1}$ leading to an estimated lower limit of the excited state lifetime in the order of tenth of picoseconds, which can be associated to the IC lifetime, in good agreement with the results reported for *p*-benzoquinone radical anion.[47]

For the Feshbach vibrational levels of the aT($D_1$-DBS), occurring above the electron detachment threshold correlating with rT($D_1$), the detachment could take place through two competing processes: a) IC as in the case of the band origin, to produce rT($D_0$); or b) vibronic coupling with the vibrational states of rT($D_1$) with $\Delta v = -1$, according to the propensity rule for vibrationally induced autodetachment.[7,30,48] The second process opens the possibility of preparing electronically excited radicals. In any case, state-selected electron kinetic energy determinations are needed to disentangle between both processes.

Once more, since the electron in a DBS has a negligible effect on the structure and vibrational frequencies of the neutral core,[7,11] the FCF for the aT($S_0$) → rT($D_1$) transition were calculated and the corresponding vibronic spectrum is shown in Figure 5 in comparison with the experimental spectrum recorded in the 3.7 eV region.

As observed in Figure 5, there is a very good agreement between both spectra. From the FCF simulation, the most intense band at 3.69 eV corresponds to the $0_0^0$ transition, followed by a group of transitions assigned to vibrational excitation of the $\nu_4 = 174$ cm$^{-1}$, $\nu_6 = 339$ cm$^{-1}$ and $\nu_8 = 523$ cm$^{-1}$ vibrational modes of rT($D_1$). These vibrational modes correspond to displacements of the $CH_3$ group associated with the geometry changes between aT($S_0$) and rT($D_1$) (Figure 5, top). In addition, the FCF for the aT($S_0$) → aT($S_3$) transition was calculated, and as shown in Figure S1 (supplementary material), there is a poor agreement between the experimental and calculated spectrum.

These results allow assigning the sharp transitions in the 3.7 eV region to the $D_1$-DBS and/or vibrational Feshbach resonances of this $D_1$-DBS.

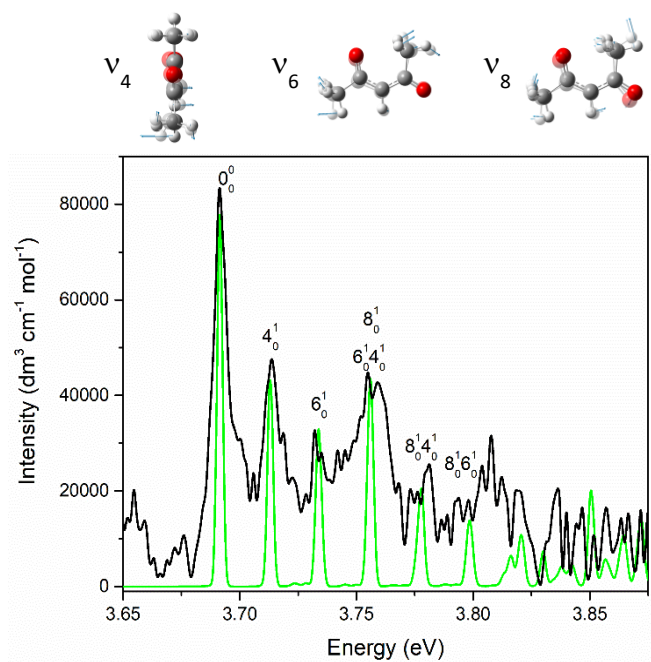

*Figure 5:* Electron photodetachment spectrum recorded in the 3.7 eV region (black). Simulated spectrum for the $S_0 \rightarrow D_1$ transition of the aT isomer at 50 K. The vibrational modes with FC activity are shown on the top.

## V. CONCLUSIONS

A single isomer of $C_5H_7O_2^-$ is populated under the experimental conditions of this work, as determined by IR spectroscopy in He-nanodroplets.

Two dipole-bound-states were identified by photodetachment spectroscopy. The first one appears in the 2.7 eV spectral region, close to the detachment threshold and corresponds to the $D_0$-DBS in which the electron in the diffused orbital is bound to the $D_0$ state of the neutral core, whose dipole moment is $\mu$ = 2.5 D. The second DBS is unexpectedly observed at 3.69 eV, ~1 eV above the electron detachment threshold, and is assigned to the $D_1$-DBS for which the excess electron is bound to the neutral core in the $D_1$ excited state with $\mu$ = 4.4 D.

We would like to note that while the present work does not report the first observation of a CE-DBS, as far as we can determine, it represents the first observation of a CE-DBS lying ~1 eV above the ADE. This result is unexpected because at this excess energy, the electron detachment is expected to be ultrafast, which would cause a very short-lived DBSs, preventing its observation. However, in the present case the lifetime of the $D_1$-DBS is limited by the internal conversion rate to the $D_0$-DBS, which requires a change in the electronic configuration of the neutral core.

The results presented here open new questions regarding the role of CE-DBS for understanding the electron capture mechanism by neutral polar molecules in their ground and excited states as well as the resonances in molecular anions that are not yet fully understood.

## VI. SUPPLEMENTARY MATERIAL

Comparison of the electron photodetachment spectrum recorded in the 3.7 eV region with the simulated spectrum for the $S_0 \rightarrow D_3$ transition.

## ACKNOWLEDGMENTS

This work has been conducted within the International Associated Laboratory LEMIR (CNRS/CONICET) and was supported by CONICET, FONCyT (PICT2019 N° 241 and PICT2021 Aplicación N° 006), SeCyT-UNC, MinCyt-Córdoba, ECOS Sud – MinCyT A22U01 and the ANR Research Grant (ANR2010BLANC040501-ESPEM, ANR17CE05000502-Wsplit and ANR-21-CE30-0004-01). The authors acknowledge access to Mendieta cluster (CCAD-UNC) and the contribution of Drs. Christophe Jouvet and Claude Dedonder-Lardeux to the experiments published here.

## AUTHOR DECLARATIONS

### Conflict of Interest

The authors have no conflicts to disclose.

### Author Contributions

R. A. Jara-Toro: Conducted all calculations (lead), Investigation (equal).

M. I. Taccone: Performed the nanodroplets-IR spectroscopy, Investigation (equal)

J. Dezalay: Investigation (supporting); Writing – review & editing (supporting).

G. von Helden: Writing – review & editing (equal).

J. A. Noble: Cold ions photodetachment spectroscopy (equal); Investigation (equal), Conceptualization (equal); Writing – review & editing (equal).

G. A. Pino: Cold ions photodetachment spectroscopy (equal); Investigation (equal), Conceptualization (equal); Writing – original draft (lead); Writing – review & editing (equal).

## DATA AVAILABILITY

The data that support the findings of this study are available from the corresponding author upon reasonable request.